\documentclass[a4,journal]{IEEEtran}
\usepackage{algorithm}
\usepackage{algorithmicx}
\usepackage{algpseudocode}
\usepackage[dvipsnames]{xcolor}
\usepackage{import}
\usepackage{tikz}
\usepackage{footnote}
\usepackage{enumitem}
\usepackage[backref=page]{hyperref}
\usepackage{ragged2e} 
\usepackage{nopageno}
\usepackage{amsthm,amsmath,amssymb}
\allowdisplaybreaks

\newtheorem{remark}{Remark}

\DeclareMathOperator*{\poly}{poly}

\hypersetup{
    colorlinks=true,
    linkcolor=blue,
    filecolor=magenta,      
    urlcolor=blue,
    pdftitle={},
    pdfpagemode=FullScreen,
    citecolor = blue,
    }
\usepackage{multirow}
\usepackage{graphicx, pifont} 
\usetikzlibrary{arrows.meta}
\usetikzlibrary{positioning,fit,backgrounds}
\usetikzlibrary{decorations.pathmorphing,patterns}

\usepackage{tikz}
\usepackage{tikz-network}
\usepackage{ifdraft}
\algnewcommand\Output{\item[\textbf{Output:}]}
\usepackage{mathtools}

\usepackage{cite}

\begin{document}



\title{ {Limitations of Fault-Tolerant Quantum Linear System Solvers for Quantum Power Flow}}

\author{{Parikshit Pareek$^\dagger$, Abhijith Jayakumar$^\star$, Carleton Coffrin$^\star$, and Sidhant Misra$^\star$ \vspace{-2em}}
\thanks{
$^\dagger$Corresponding Author and is currently with Department of Electrical Engineering, Indian Institute of Technology Roorkee. He was previously with Los Alamos National Laboratory, NM, USA where this work was done.\\
$^\star$ are currently with Los Alamos National Laboratory, NM, USA.\\
 pareek@ee.iitr.ac.in; abhijithj@lanl.gov; cjc@lanl.gov, sidhant@lanl.gov. \\
The authors acknowledge the funding provided by LANL’s Directed Research and Development (LDRD) project: ``High-Performance Artificial Intelligence"(20230771DI) and the Department of Energy (DOE), USA under the Advanced Grid Modeling (AGM) program. 
The research work conducted at Los Alamos National Laboratory is done under the auspices of the National Nuclear Security Administration of the U.S. Department of Energy under Contract No. 89233218CNA000001}}

\maketitle

\begin{abstract}
Quantum computers hold promise for solving problems intractable for classical computers, especially those with high time or space complexity.  {Practical \emph{quantum advantage} can be said to exist for such problems when the end-to-end time for solving such a problem using a classical algorithm exceeds that required by a quantum algorithm}. Reducing the power flow (PF) problem into a linear system of equations allows for the formulation of quantum PF (QPF) algorithms, which are based on solving methods for quantum linear systems such as the Harrow-Hassidim-Lloyd (HHL) algorithm. Speedup from using QPF algorithms is often claimed to be exponential when compared to classical PF solved by state-of-the-art algorithms. We investigate the potential for practical quantum advantage in solving QPF compared to classical methods on gate-based quantum computers. \textcolor{black}{Notably, this paper does not present a new QPF solving algorithm but} scrutinizes the end-to-end complexity of the QPF approach, providing a nuanced evaluation of the purported quantum speedup in this problem. Our analysis establishes a best-case bound for the HHL-based quantum power flow complexity, conclusively demonstrating that the HHL-based method has higher runtime complexity compared to the classical algorithm for solving the direct current power flow (DCPF) and fast decoupled load flow (FDLF) problem.  {Notably, our analysis and conclusions can be extended to any quantum linear system solver with rigorous performance guarantees, based on the known complexity lower bounds for this problem.}  Additionally, we establish that for potential practical quantum advantage (PQA) to exist it is necessary to consider DCPF-type problems with a very narrow range of condition number values and readout requirements. 
\end{abstract}
\begin{IEEEkeywords}
Quantum Computing, Direct Current Power Flow (DCPF), Fast Decoupled Load Flow (FDLF)
\end{IEEEkeywords}

\section{Introduction}
Providing electrical energy to the world and fighting climate change by switching from fossil fuel--powered to electric-powered transportation will require electricity generation to increase by more than two and half times by 2050 from 2021 levels \cite{iea2021netzero}. This increased reliance on electricity will make secure power grid operation an even more important challenge. Modern power system operation relies heavily on the ability of system operators to accurately and efficiently compute the system state. The fundamental computational tool underpinning this analysis is the power flow problem, which is solved to obtain node voltages and line flows for a given set of nodal power injections \cite{john1994power}. Direct current power flow (DCPF) approximation is a popular approach as it provides a fast, unique, and reliable solution to the power flow problem  \cite{dcpf,lossyDC,miso,BPM002_D,BPM_002,BPM_002_B}. From a computational standpoint, the DCPF problem translates to solving a linear system of equations given a nodal power injection vector. Therefore, an increase in system size and changes to the injection vector due to uncertainty increases the computational effort required to conduct DCPF analysis. Another prominent linear power flow formulation, fast decoupled load flow (FDLF), also reduces to solving a sequence of linear systems of equations \cite{john1994power}. \textcolor{black}{Solving fast DCPF and FDLF is imperative for secure power system operations under uncertainty, as they serve as essential subroutines in various probabilistic and stochastic analyses, such as probabilistic power flows. For instance, evaluating the percentage of lines violating power flow constraints over one million possible load points leads to a total runtime of $10^6 \times \text{Single-instance Runtime}$. In such cases, even a marginal speedup in the runtime of a single instance of DCPF or FDLF can have a significant impact on the overall computational efficiency, especially when considering the large number of load samples involved, even with considerable parallelization.} In practice, algorithms such as the conjugate gradient (CG) method are used to solve these linear power flow problems for large-scale systems, exploiting features such as sparsity for faster convergence to the solution \cite{shewchuk1994introduction}.

Quantum computing (QC), in theory, can provide provably asymptotic speedups over classical computers for many problems of practical interest 
\cite{kockum2023lecture,abhijith2022quantum,dalzell2023quantum}. QC achieves this by exploiting inherently quantum properties such as superposition and entanglement. This leads to novel algorithmic techniques in QC that cannot be emulated efficiently by classical computers. One such algorithmic primitive developed in QC literature is an exponentially fast linear system solver, first introduced in the work of Harrow, Hassidim, and Lloyd (HHL) \cite{harrow2009quantum} and improved on by others \cite{ambainis2010variable, childs2017quantum}. Exponential speedup in these techniques comes from the fact that a linear system solving subroutine can solve a sparse linear system in a time that scales only logarithmically with system size, if subroutine input is provided as a quantum state and if relevant output information can be extracted using a small number of quantum measurements. The input/output model used in these algorithms is markedly different from those used in classical computing. This suggests that, in general, a true end-to-end exponential speedup cannot be claimed over classical algorithms. This point has been noted in QC literature \cite{aaronson2015read}, and in some cases the claimed exponential speedup has been shown to vanish when the classical computer is given access to a data input/output model that appropriately matches the data model used by the QC algorithm \cite{tang2019quantum,tang2021quantum, chia2022sampling}.

\begin{figure*}[t]
    \centering
    \resizebox{0.3\textwidth}{3.2cm}{
     \begin{tikzpicture}
    \vspace{-2em}
    \node[circle, draw, minimum size=0.2pt,fill=black!10] (node1) at (0,0) {1};
    \node[circle, draw,fill=black!10, minimum size=0.2pt] (node2) at (2,0) {2};
    \node[circle, draw,fill=black!10, minimum size=0.2pt] (node3) at (2,-2) {3};
    \node[circle, draw,fill=black!10, minimum size=0.2pt] (node4) at (0,-2) {4};

   \node at (node1) [above, yshift=4mm, text=red] {$v_1$\textcolor{black}{$\angle0$}};
    \node at (node2) [above,yshift=4mm, text=red] {$v_2\angle\theta_1$};
    \node at (node3) [below,yshift=-4mm, text=red] {$v_3\angle\theta_3$};
    \node at (node4) [below, yshift=-4mm, text=red] {$v_4\angle\theta_4$};
    \node[circle,fill=green!40, draw, minimum size =0.4pt] (g2) at (3.5,0) {\footnotesize \textcolor{black}{$L_1$}};
      \node[circle,fill=green!40, draw, minimum size =0.4pt] (g1) at (-1.5,0) {\footnotesize \textcolor{black}{$G_1$}};
        \node[circle,fill=green!40, draw, minimum size = .3pt] (g4) at (-1.5,-2) {\footnotesize \textcolor{black}{$G_2$}};
    \draw[->]  (g4) -- (node4) node[midway, above, text=black] {\footnotesize$P_4$};
    \draw[->] (node2) -- (g2) node[midway, above, text=black] {\footnotesize$P_2$};
   \draw[->] (g1) -- (node1) node[midway, above, text=black] {\footnotesize$P_1$};
    \node[circle,fill=green!40, draw,minimum size = .4pt] (l) at (3.5,-2) {\footnotesize \textcolor{black}{$L_3$}};
    \draw[->] (node3) -- (l) node[midway, above, text=black] {\footnotesize$P_3$};
   \draw[draw,-,thick] (node1) -- (0.7,0);
    \draw[decoration={aspect=0.3, segment length=2mm, amplitude=1.5mm,coil},decorate] (0.7,0) -- (1.3,0) node[midway, above,text=black,yshift = 1mm] {\footnotesize$b_{12}$};
    \draw[draw,-,thick] (1.3,0) -- (node2);
    \draw[draw,-,thick] (node2) -- (2,-0.7);
    \draw[decoration={aspect=0.3, segment length=2mm, amplitude=1.5mm,coil},decorate] (2,-0.7) -- (2,-1.3) node[midway, right,text=black,xshift = 1mm] {\footnotesize$b_{23}$};
    \draw[draw,-,thick] (2,-1.3) -- (node3);
    \draw[draw,-,thick] (node1) -- (0.7,-0.7);
    \draw[decoration={aspect=0.3, segment length=2mm, amplitude=1.5mm,coil},decorate] (0.7,-0.7) -- (1.3,-1.3) node[midway, above,text=black,yshift=2mm,xshift = 2mm] {\footnotesize$b_{13}$};
    \draw[draw,-,thick] (1.3,-1.3) -- (node3);
    \draw[draw,-,thick] (node1) -- (0,-0.7);
    \draw[decoration={aspect=0.3, segment length=2mm, amplitude=1.5mm,coil},decorate] (0,-0.7) -- (0,-1.3) node[midway, left,text=black,xshift=-1mm] {\footnotesize$b_{14}$};
    \draw[draw,-,thick] (0,-1.3) -- (node4);
    \draw[draw,-,thick] (node4) -- (0.7,-2);
    \draw[decoration={aspect=0.3, segment length=2mm, amplitude=1.5mm,coil},decorate] (0.7,-2) -- (1.3,-2) node[midway, below,text=black,yshift=-1mm] {\footnotesize$b_{34}$};
    \draw[draw,-,thick] (1.3,-2) -- (node3);
   \node[minimum size = .4pt] (v) at (-2,-1) {\small \textcolor{black}{$v_i = 1 $ (pu) $\forall i$}};
\end{tikzpicture}}
     \hfill
\includegraphics[width=0.35\textwidth]{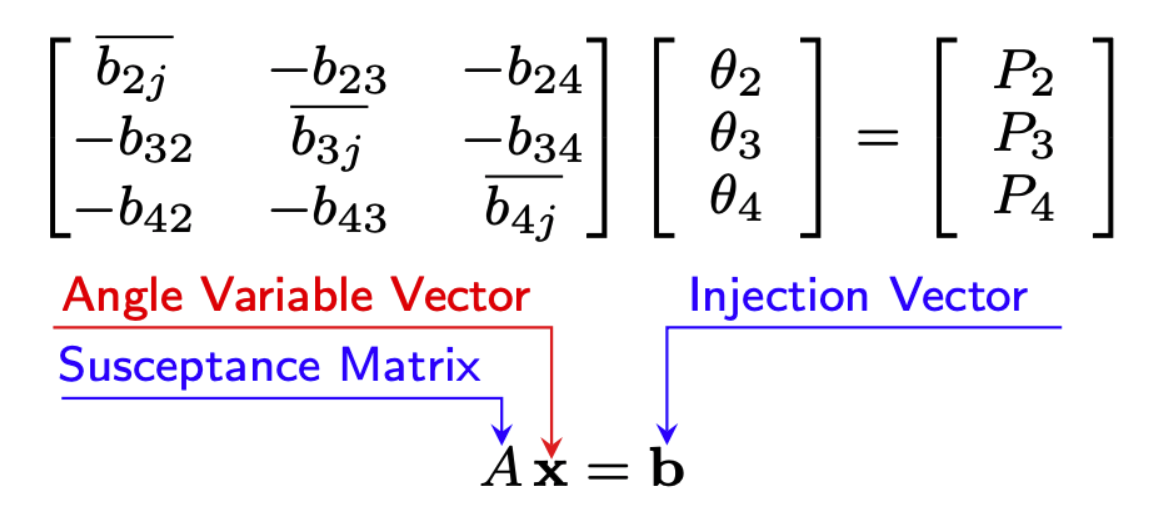}
    \hfill   
    \hspace{-1em}\includegraphics[width=0.5\columnwidth]{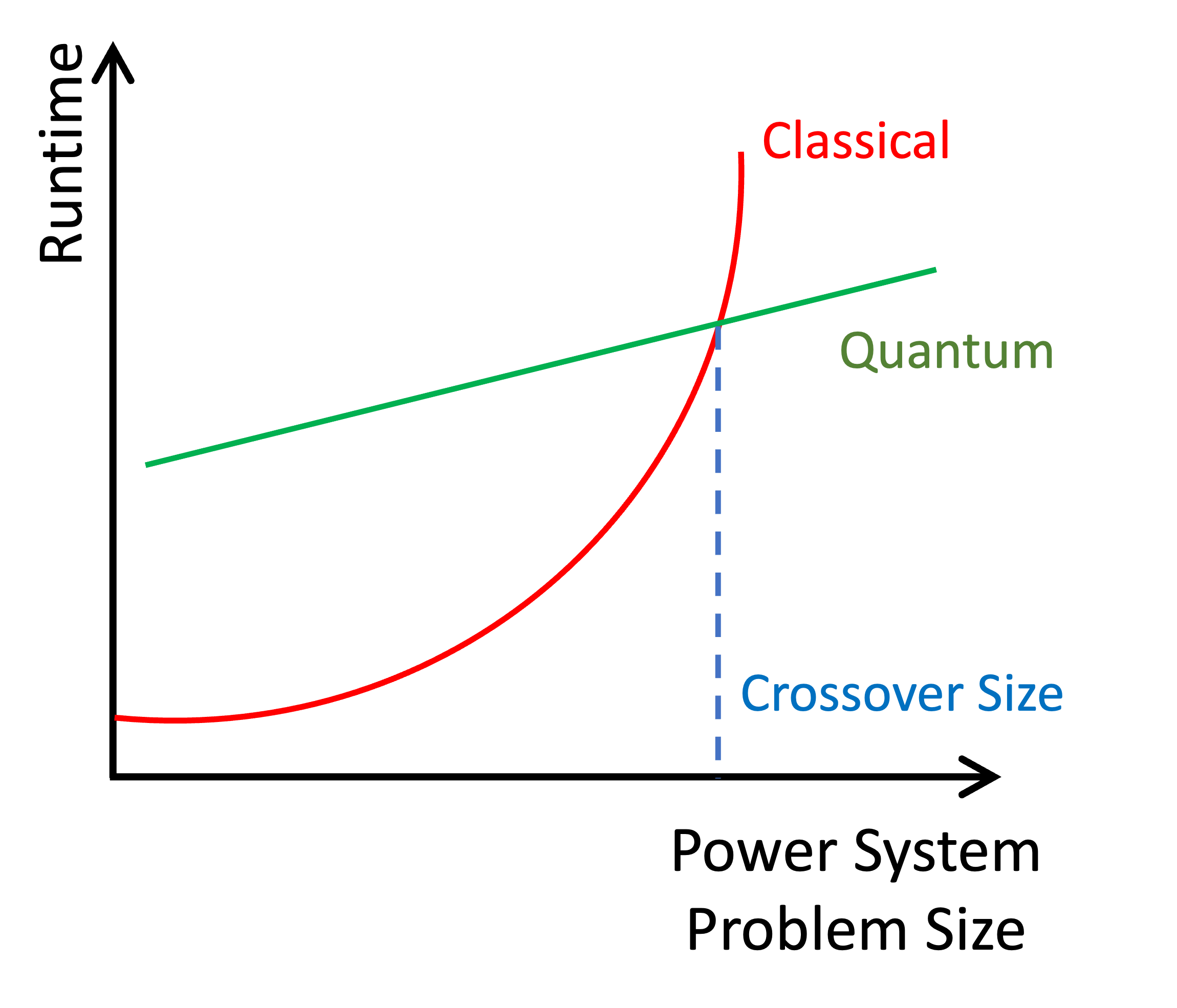}
    \caption{Power network data, equivalent linear system of equations, and the idea of practical quantum advantage (PQA). \textbf{Left:} Consider a simplified power grid with four ``buses" (nodes) connected by five ``branches" (lines). Each line has a certain ``susceptance" that affects how easily electricity flows through it. Using direct current approximations \cite{dcpf}, we can approximate the network by focusing only on the susceptances and ignoring other electrical properties (in the DC power flow approximation, the network is assumed to be lossless, the phase-angle differences between nodes are considered small, and all node voltages are assumed to be approximately equal to one per unit \cite{dcpf}). The unknowns in this system are the bus angle values (shown in red), which depend on known quantities like susceptance (shown in black). Without any loss of generality, we can set one of the bus angle values to zero. \textbf{Center:} To solve the problem, DCPF converts the grid into a set of equations. It uses the susceptance values, known bus angles, and power injections (how much electricity enters or leaves each bus) to calculate the unknown bus angle values. Diagonal dominance arguments show that $A$ in this system is always positive \textcolor{black}{definite \cite{dcpf,horn2012matrix}, and one can always construct a $A^\dagger A\mathbf{x} = A^\dagger \mathbf{b}$, where $A^\dagger$ is complex conjugate of matrix $A$, and $A^\dagger A$ is Hermitian}. \iffalse Notice the diagonal dominance of the matrix as $\overline{b_{kj}} = \sum^N_{j=1} b_{kj}$ \cite{dcpf}. \fi \textbf{Right:} Complexity analysis of classical and quantum approaches to this problem tells us how the respective runtimes of these approaches scale with problem size and other parameters. PQA can exist only if there is a crossover point as depicted in this figure. Such a crossover implies that there exists a system size beyond which quantum algorithms can be faster for solving this problem.}
    \label{fig:qaa}
    \vspace{-2em}
\end{figure*}

\begin{figure*}[t]
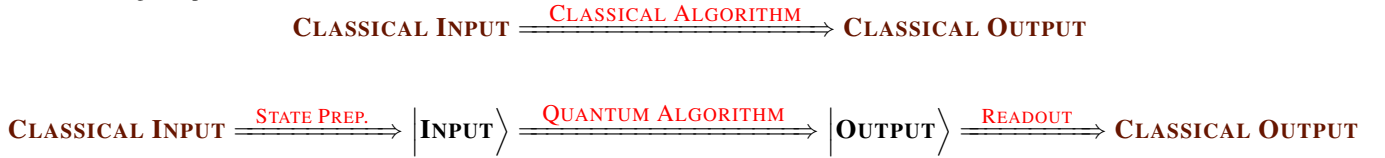

    \centering
   \begin{align*}
   \textcolor{Sepia}{\textsc{\textbf{Classical Input}}}  & \xRightarrow{  \scalebox{0.9}{~~ \textsc{\textcolor{red}{Classical Algorithm}}}~~ } \textcolor{Sepia}{\textsc{\textbf{Classical Output}}} \\ \\
   \textcolor{Sepia}{\textsc{\textbf{Classical Input}}} \xRightarrow{\scalebox{0.8}{\textsc{~~ \textcolor{red}{State Prep.}}}~~ } \textcolor{black}{\Big |\textsc{\textbf{Input}}\Big \rangle} &\xRightarrow{\scalebox{0.9}{\textsc{~~ \textcolor{red}{Quantum Algorithm}}}~~ } \textcolor{black}{\Big |\textsc{\textbf{Output}}\Big \rangle} \xRightarrow{\scalebox{0.8}{\textsc{~~ \textcolor{red}{Readout}}}~~ } \textcolor{Sepia}{\textsc{\textbf{Classical Output}}}
    \end{align*}
    \caption{A simple information flow schematic of solving a problem using classical and quantum algorithms. From Figure \ref{fig:qaa}, classical input is $\mathbf{b}$, and classical output is $\mathbf{x}$. The quantum algorithm encodes the matrix $A$ (considering $A$ as a Hamiltonian for a quantum system) in terms of the time evolution operator $e^{iAt}$. As depicted, solving a linear system using quantum computing by means of the HHL algorithm involves three steps \cite{schuld2018supervised}: (i) state preparation; (ii) state propagation (quantum algorithm), and (iii) readout. In the state preparation step problem, parameter $\mathbf{b}$ is encoded into a quantum state $|\Psi_o\rangle$. Following this, the second step is the application of quantum gates (unitary operations) to evolve the initial quantum state $|\Psi_o\rangle$ into the final state $|\Tilde{\textbf{x}}\rangle$ based on the properties of the matrix $A$. Here, $|\Tilde{\textbf{x}}\rangle$ is a quantum state and the third step is a readout, where $|\Tilde{\textbf{x}}\rangle$ is measured to extract the solution to the linear system $\widehat{\mathbf{x}}$.
    We assess the quantum advantage based on the end-to-end runtime complexities of the DCPF problem. Here, end-to-end complexity in QC refers to the comprehensive time complexity from input to output, encompassing all stages of solving the DCPF problem and not just the cost of running a quantum circuit that is a subroutine of the full solution method. 
    The end-to-end complexity determines the existence of PQA for these problems by showing how the total runtime required to solve a problem scales with system size and other parameters in realistic DCPF problems. }
    \label{fig:outline}
    \vspace{-1.5em}
 \end{figure*}

Recently, QC has garnered significant attention in the power systems community, with efforts directed towards solving power flow problems through QC techniques, referred to as quantum power flow (QPF) \cite{qpf,9122420,saevarsson2022quantum,liu2022quantum,golestan2023quantum,vereno2023exploiting,kaseb2023quantum,gao2023solving,amani2023quantum,feng2023noise,neufeld2023hybrid}. The primary rationale behind these endeavors lies in QC's theoretical ability to efficiently solve a linear systems of equations. This is particularly pertinent as various power flow formulations, such as DCPF and FDLF, are either equivalent to or can be reduced to solving linear systems of equations \cite{dcpf,john1994power}. In \cite{qpf, saevarsson2022quantum, 9122420}, the authors suggest employing the HHL algorithm to solve FDLF on a quantum computer, implying a potential exponential speedup due to the runtime complexity of HHL. Authors in \cite{saevarsson2022quantum} present results obtained using a real quantum computer for solving the FDLF problem. The authors in \cite{qpf} and \cite{9122420} propose exponential speedup, but the complexities associated with the readout of FDLF iteration outputs are not addressed. This aspect, as highlighted in \cite{saevarsson2022quantum}, is crucial for the execution of FDLF. Additionally, in \cite{saevarsson2022quantum}, the authors rightly argue that quantum memory is indispensable for attaining quantum speedup in FDLF and suggest probabilistic Monte-Carlo simulation-type problems as potential applications of QPF. However, the paper does not analyze end-to-end complexity, which includes stages like state preparation. To alleviate qubit requirements, \cite{gao2023solving} proposes a hybrid method. Additionally, \cite{feng2023noise,liu2022quantum} describe variational QPF to reduce the circuit depth needed to solve QPF compared to that needed for HHL-based QPF \cite{qpf}. Authors in \cite{liu2022quantum} provide a comprehensive review of different QPF formulations and highlight the need to multiply the `downloading and uploading' complexities in HHL-QPF complexity. However, the effect of these complexities on quantum speedup remains to be explored.

Overall a through assessment of the requirements for quantum advantage is lacking in the current literature. Quantum advantage refers to the potential computational superiority of quantum computers over classical computers for specific problems \cite{preskill2012quantum}. The challenge of a PQA assessment can be shown by plotting a runtime complexity or resource requirement curve, as illustrated in the rightmost subplot of Figure  \ref{fig:qaa}.
In the context of DCPF, identifying the crossover point on this plot, that is, where the quantum computer's resource requirements become more favorable than the best-known classical method, holds significant implications for the practical adoption of QC in power systems. If such a crossover point does not exist or if the crossover point exceeds the size of real-world problems (e.g., a billion-node power flow problem), the adoption of QC may not be practically justified. For the QPF problem in specific, achieving PQA involves demonstrating that a quantum algorithm can efficiently provide the complete network state vector (including both the voltage angle and magnitude, depending on the formulation) by solving a power flow problem significantly faster or with fewer computational resources than and with the desired level of accuracy of the best-known classical algorithms.


In this paper, we elaborate on the complexities associated with each step of solving the QPF, shown in Figure \ref{fig:outline}. Further, we critically weigh the exponential speedup claim of HHL-QPF methods against various HHL caveats \cite{aaronson2015read}. 
Further, by completing an end-to-end complexity assessment for this problem, we attempt to build a framework to assess the following:
\begin{itemize}
    \item \textit{Does practical quantum advantage exist in `solving' the DCPF problem?}
    \begin{itemize}
        \item \textit{If yes, at what power system size will the runtime complexity of quantum computing methods surpass that of classical methods?}
        \item \textit{If no, what essential features must a DCPF-type problem have to show potential for PQA?}
    \end{itemize}
\end{itemize}

\textcolor{black}{It is important to highlight that this work does not provide a new QPF solving algorithm}, but instead hones in on the runtime complexity of QPF by considering end-to-end time requirements and incorporating key parameters from realistic AC power network datasets \cite{pglib}. The main contributions of this paper can be summarized as follows: 
\begin{itemize}
    \item Demonstrating that the end-to-end runtime complexity of solving the DCPF problem using the HHL-QPF algorithm is asymptotically worse than that of the classical CG method. When considering the readout \cite{cramer2010efficient} requirement in QC and condition number growth with network size ($N$ Buses) for realistic AC power network datasets \cite{pglib}, the complexity of HHL-QPF becomes $\mathcal{O}(N^{4.62}\log N)$, whereas the CG method's complexity for solving an $N$-Bus DCPF problem remains $\mathcal{O}(N^{1.90} \log N)$. The effect of runtime overhead of state preparation is also discussed. 
    \item Identifying the power network parameter range (condition number) and readout level\footnote{Readout level refers to the amount of classical information needed to extract the desired result from a quantum computation. For example, if we only need the mean of all states, the readout level is one, as we only require a single classical value. Consider a power system with N=100 buses. If we're interested in the angles of buses 1, 20, 33, 55, and 99 in a 100-bus system, the readout level would be five, even though the full vector $\mathbf{x}$ contains angles for 99 buses.} for the power flow--type problems for which PQA can potentially exist. We show that for readout requirements of $D$ states with condition number $\kappa$, problem parameters must satisfy\footnote{ Where, condition number $\kappa$ is the ratio of largest to smallest eigenvalue and $s$ is the sparsity of matrix $A$ defined as largest number of non-zero elements in a row.}  $D\kappa^{3/2} \ll Ns^{-1}$ to have any potential PQA using the HHL algorithm.
\end{itemize}


{\color{black}
\begin{remark}
Unlike HHL and its improvements, the Variational Quantum Linear Solver (VQLS) is a hybrid quantum-classical algorithm \cite{bravo2023variational}, where  ansatz is updated using a hybrid quantum-classical optimization loop, to achieve local or global cost below user-specified threshold. This makes VQLS a heuristic algorithm, unlike HHL, which is a purely quantum algorithm with the stopping criterion defined by the measurement of an ancilla bit (Appendix \ref{app:hhl} discusses this criteria of algorithm success) \cite{harrow2009quantum,stepbystepHHL}. Therefore, the end-to-end runtime complexity discussion does not apply to VQLS. Furthermore, due to its heuristic nature, an empirical study is required to evaluate the effectiveness of VQLS in solving DCPF (DC Power Flow) problems and to understand its practical applicability. It is worth noting that the paper introducing VQLS \cite{bravo2023variational} only conducts such a study to provide an upper bound on the quality of the solution, not on the runtime. 
\end{remark}}

\begin{remark}
 {We only consider quantum linear system solvers designed to run on a error corrected, fault tolerant quantum computers. Most theoretical claims of quantum advantage are made for this type of quantum computers. While quantum algorithms can be run on Noisy Intermediate Scale Quantum (NISQ) devices, these do not come with rigorous performance guarantees and can be limited in terms of scalability \cite{chen2023complexity, takagi2022fundamental}. }
    
\end{remark}

\section{DC Power Flow Problem}
The DCPF equations can be obtained by applying several simplifying assumptions to the non-linear alternating current (AC) branch flow equations \cite{dcpf}. These assumptions lead to conditions in which the power balance equations can be expressed as a weighted sum of the angle differences between buses. The weights in this sum are represented by the elements of the susceptance matrix as shown in Figure \ref{fig:qaa}. In matrix form, the DCPF is
\begin{align}\label{eq:linsys}
    A \mathbf{x}=  \mathbf{b}
\end{align}
with $A$ being positive definite, symmetric matrix of size $N \times N$ where $N$ is one less than system size, and $\mathbf{x}$ and $\mathbf{b}$ are $N$-dimensional vectors.\footnote{We have opted to use the standard linear system notation $A \mathbf{x} = \mathbf{b}$ instead of the standard DCPF notation $B \boldsymbol{\theta} = \mathbf{p}$. Here, $B$ is the susceptance matrix, $\boldsymbol{\theta}$ is the node phase-angle vector, and $\mathbf{p}$ is the nodal real-power injection vector. This choice facilitates compatibility with the analysis of linear system solving complexity for both classical and quantum paradigms.}

\begin{figure*}
    \centering
    \includegraphics[width=\textwidth]{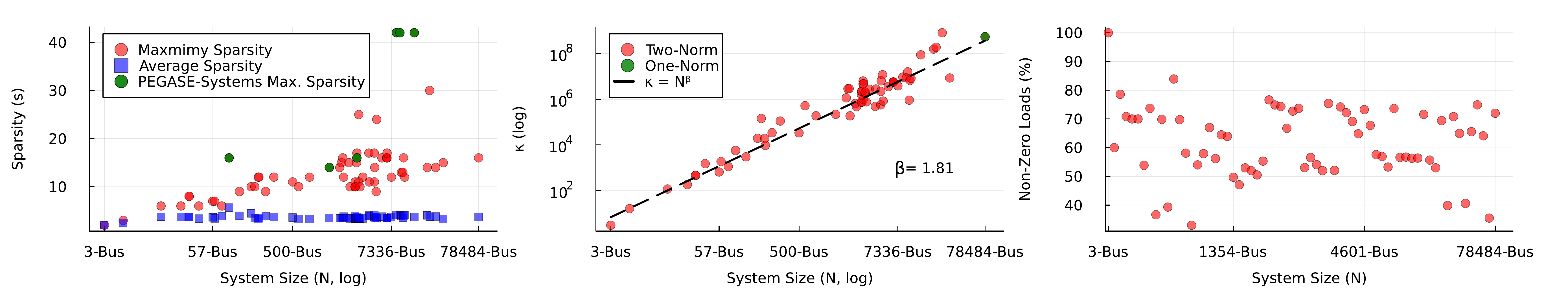}
    \caption{Network properties of the \texttt{PGLib} \cite{pglib} power transmission network datasets reveal structural insights and potential numerical challenges in solving DCPF. \textbf{Left:} Level of sparsity ($s$) for different power system size ($N$-buses). Here, average sparsity is the average number of non-zero elements in any row of matrix $A$. The PEGASE grids tend to have higher $s$ values due to subsystem equivalencies. Sparsity levels are below 30 for most of the 66 power grids from \texttt{PGLib}, with an average sparsity under 10, highlighting the potential for efficient sparse matrix techniques. \textbf{Center:} Polynomial relationship of condition number ($\kappa$) of $A$ with power system size ($N$ buses). The one-norm  is used to lower bound $\kappa$ for the 78484-Bus system due to computational restrictions of calculating the two-norm of such a large system. Condition number exhibits a polynomial (near-quadratic) growth trend with increasing system size ($N$-buses), signaling potential numerical challenges and runtime increase in solving DCPF. \textbf{Right:} Percentage of non-zero elements in injection vectors (sparsity percentage) for different \texttt{PGLib} systems. It shows that injection vector ($\mathbf{b}$ in DCPF formulation) sparsity will not be able to reduce the encoding complexity of $\mathbf{b}$ vector beyond a certain level. More importantly, due to the nature of power grid operation applications, one cannot make strong assumptions about or requirements on load-vector sparsity levels. \textcolor{black}{We use \texttt{PowerModels.jl}\cite{8442948} to construct the susceptance matrix for all these \texttt{PGLib} \cite{pglib} systems.}}
    \label{fig:system_details}
\end{figure*}

The runtime complexity to solve \eqref{eq:linsys} using Gaussian elimination is $\mathcal{O}(N^3)$, or, precisely, $\mathcal{O}(N^\omega)$, where $\omega \in [2,2.372)$ denotes the matrix multiplication exponent \cite{dalzell2023quantum}. This complexity becomes prohibitive when solving DCPF for large-scale systems because an industry scale network of 10,000 buses will require $\sim 10^{10}$ operations, which is daunting for classical computers processing operations at gigahertz speeds. Further, exact methods  {matrix inversion (e.g. Gaussian Elimination)} are not always able to exploit the sparsity present in system matrix $A$. In a power network, each node is connected to only a few other nodes of the network grid. Thus, sparsity is present in $A$ for all practical power networks. Figure   \ref{fig:system_details} (left) shows the sparsity level for different systems from \texttt{PGLib}\cite{pglib} library, which includes over sixty open-access transmission network datasets.


Moreover, the computational efficiency can be improved by not necessitating the precise solution of $\mathbf{x}$ in the DCPF problem. Typically, errors within the range of $10^{-4}$ to $10^{-6}$ are considered reasonable, varying based on the specific application requirements. This range is deemed acceptable due to the resolution limitations of physical measurement devices within a power network. The impact of the error value on runtime is quantified by the parameter $\varepsilon$ in our complexity discussions.

\section{Classical Method Complexity} 
The conjugate gradient (CG) method is among the best algorithms in terms of asymptotic complexity for solving linear systems of equations with a sparse, positive definite matrix $A$. As the power system matrices fulfill both conditions, we focus on analyzing CG as one of the best ways to classically solve DCPF. We define $s$ as the maximum number of non-zero values in a row, i.e., sparsity, and $\kappa$ as condition number ($\kappa= \|A^{-1}\|\|A\|$). The total complexity of CG can be split into two parts: complexity of dominant operations and number of times dominant operations are used. 

Considering the dominant operation complexity of matrix-vector multiplication, $\mathcal{O}(Ns),$ and an upper bound on number of iterations required to achieve the error $\varepsilon_c$ (details are given in appendix), the overall complexity of the CG method for solving \eqref{eq:linsys} is $\mathcal{O}(N\sqrt{\kappa}s\log({1}/{\varepsilon_c}))$. Further, read-write complexity for the sparse matrix $A$ is $\mathcal{O}(Ns)$ (higher than the $\mathbf{b},\mathbf{x}$ readout cost, which is $\mathcal{O}(N)$), and this additional cost is additive to the runtime complexity. Here, additive means that we read $A,\mathbf{b}$ once before solving \eqref{eq:linsys} and read $\mathbf{x} \in \mathbb{R}^N$ once after solving the problem. To obtain $\mathbf{x}$, CG does not need to be repeated but only needs to be read with complexity $\mathcal{O}(N)$. This is important as additive terms are subsumed into dominant complexity terms and do not alter the end-to-end complexity. Thus, the end-to-end complexity of solving DCPF \eqref{eq:linsys} using CG is $\mathcal{O}(Ns + N\sqrt{\kappa}s\log({1}/{\varepsilon_c}) + N) = \mathcal{O}(N\sqrt{\kappa}s\log({1}/{\varepsilon_c}))$.\footnote{For basic Big-O notation introduction, readers can refer to these notes: \href{https://web.mit.edu/16.070/www/lecture/big_o.pdf}{Link}.} Along with exploiting sparsity, CG runtime complexity has a square root dependence condition number of $\kappa$, making it highly efficient for DCPF as $\kappa$ has polynomial growth, as shown in Figure \ref{fig:system_details} (center). 

From the definition of $\varepsilon_c$ and error vectors ($\mathbf{e}^o$ and $\mathbf{e}^i$; details are given in appendix), it is clear that the CG error upon convergence will depend on the initial point $\mathbf{x}_o$. If a starting point is closer to the final solution, the same value of $\varepsilon_c$ can be achieved in fewer CG iterations. This feature of CG has led to considerable efforts towards finding `good' starting points of phase-angles ($\mathbf{x}$) in DCPF problems, particularly when DCPF needs to be solved repeatedly with small perturbations in power injections ($\mathbf{b}$), known as DCPF hot-starting \cite{dcpf}. The error is expressed in terms of the energy norm, $\| e^i \|_{A}$ (see appendix for more details on error definitions.).\footnote{Note that $\lambda_1||e||_2\leq \|e^i\|_{A} = \sqrt{(e^i)^T A e^i} \leq \lambda_N\|e^i\|_2$ if $A$ is PSD where $\lambda_1 < \dots < \lambda_N$ is the ordered set of eigenvalues of $A$. For sparse matrices, this maximum eigenvalue can be often bounded by a constant. Moreover, for multiple power systems, the susceptance matrix has eigenvalues less than one (30, 118, 200, 500, 1354, and 2383-Bus system from \cite{pglib}). More details on effect of error definition on CG complexity are given in appendix.} Therefore, the effects of the definition of $\varepsilon_c$ on runtime complexity must be considered carefully when comparing CG to any other classical or quantum algorithm for solving \eqref{eq:linsys}.

\section{Quantum Advantage Analysis of HHL-QPF}
We now examine the complexity of different components of HHL-QPF.\footnote{Although the discussion presented in this section is centered around using HHL for QPF, our analysis in general applies to other quantum linear solvers as well.  Table \ref{tab:complexity} shows the complexity of another quantum algorithm to solve a linear system of equations called VTAA-HHL \cite{ambainis2010variable}. Complexities of state-of-the-art quantum algorithms for linear systems show linear scaling with condition number. They are also subject to the same state preparation and read-out considerations that we discuss in this work \cite{lin2020optimal,dalzell2023quantum}.}
The objective is to systematically analyze the complexity of the operations employed in solving QPF, culminating in an understanding of the end-to-end complexity of solving the DCPF problem.

\subsection{Reading the Problem: State Preparation Complexity}
To solve DCPF using a quantum computer, the problem must first be represented in terms of quantum states. A vector of $N$ elements can be represented using $n=\lceil\log(N)\rceil$ qubits, achieving seemingly exponential compression in memory \cite{kockum2023lecture,abhijith2022quantum}. The process of transferring data from classical memory to a quantum computer is referred to as state preparation or encoding. Quantum state preparation is accomplished through the manipulation of qubits, converting classical information into a quantum superposition. The choice of encoding method depends on the particular quantum algorithm (such as HHL in the case of QPF) used to solve the problem. The encoding type is also closely tied to the performance of the particular algorithm. In the context of HHL, the technique employed for preparing a quantum state with information from vector $\mathbf{b}$ is known as \textit{amplitude encoding}. In the \textit{amplitude encoding} technique, data is encoded into the amplitudes of a quantum state. A normalized classical $N$ element power injection vector $\mathbf{b}$ is represented by the amplitudes of an $n$-qubit quantum state $\left|\mathbf{b}
\right\rangle$ as $\left|\mathbf{b}\right\rangle=\sum_{i=1}^N b 
_i|i\rangle$. 
Here, $b_i$ is the $i$-th element of normalized injection vector $b_{norm}$ (normalized injection at $i$-th bus), and $|i\rangle$ is the $i$-th computational basis state \cite{schuld2018supervised,stepbystepHHL}. 

The QPF problem demands the ability to perform arbitrary state preparation for \textit{amplitude embedding}. This necessity arises because the injection vector $\mathbf{b}$ lacks a specific structure, and individual injection values are represented as arbitrary floating-point numbers. Further, to prepare an arbitrary $n$-qubit state, at least $N = 2^n$ elementary quantum gates are required, i.e., gate complexity is lower bounded by $\Omega(N)$ \cite{dalzell2023quantum,plesch2011quantum}. Assuming that each unitary gate operation (one or two-qubit gate) takes unit time, the runtime complexity to prepare an arbitrary quantum state $\mathcal{O}(T_b)$  for the injection vector is $\mathcal{O}(T_b) = \mathcal{O}(N)$\cite{dalzell2023quantum}.

An inventive method for accessing arbitrary classical data in a QC is to use a theorized quantum random access memory (QRAM), allowing for coherent access to classical information. The proposal to introduce and employ QRAM addresses challenges linked to state preparation or limitations in inputting data in quantum linear algebra applications \cite{aaronson2015read, dalzell2023quantum}. With those caveats on QRAM's practicality, the circuit depth (and runtime complexity) for accessing data from QRAM is demonstrated to be $\mathcal{O}(T_b) = \mathcal{O}(\log(N))$ \cite{dalzell2023quantum}. However, there is a one-time cost associated with initially loading QRAM with data. Moreover, in practice, QRAM may also cause a runtime increase due to error correction and fault tolerance overheads \cite{di2020fault}. Further, as shown in Figure  \ref{fig:system_details} (right), the load vector is not sparse in nature and the majority of \texttt{PGLib} systems have 40-80\% non-zero load values. Therefore, exploiting sparsity to encode $\mathbf{b}$ will not be  beneficial.

\subsection{Solving the Problem: State Propagation Complexity}
The quantum mechanics equivalent of solving linear system \eqref{eq:linsys} is to construct $|\mathbf{x} \rangle = A^{-1}|\mathbf{b}\rangle$. To achieve this, HHL relies on the fact that if $A$ is Hermitian, i.e., $A = A^\dagger$, then $A$ can be considered a Hamiltonian for a quantum system, and the time evolution operator of such a quantum system is given by $e^{iAt}$ \cite{harrow2009quantum}. Other stages involved in the solution process include quantum phase estimation, ancilla bit rotation, and inverse quantum phase estimation.

The quantum simulation complexity in HHL \cite{harrow2009quantum} is given by $\mathcal{O}(t_os^2\log(N))$, where $t_o = \mathcal{O}(\kappa/\varepsilon_h)$ is chosen to achieve the final error $\varepsilon_h$. Consequently, quantum simulation can be executed in $\mathcal{O}(\log(N) \kappa s^2/\varepsilon_h)$ time. In HHL, it is necessary to repeat this solution procedure to attain the ancilla qubit as $|1\rangle$ with high probability. This requires $\mathcal{O}(\kappa)$ repetitions with \textit{amplitude amplification} \cite{brassard2002quantum} to boost the final success probability of the algorithm. Combining these terms, the overall runtime complexity of HHL, given an initial vector state $|b\rangle$, is $\mathcal{O}(\kappa^2s^2\log(N)/\varepsilon_h)$ \cite{harrow2009quantum}. However, as mentioned earlier, a quantum state for the injection vector $\mathbf{b}$ needs preparation, introducing $\mathcal{O}(T_b)$ complexity. Additionally, this state must be prepared for each $\mathcal{O}(\kappa)$ repetition required in the amplitude amplification. This is because once a quantum state is measured, it is destroyed (a phenomenon known as wave function collapse in quantum mechanics) \cite{kockum2023lecture}. Therefore, the overall complexity of HHL, considering state preparation, is $\mathcal{O}(\kappa(T_b + \kappa s^2\log(N)/\varepsilon_h))$, where $T_b$ either refers to the complexity of QRAM or that of a circuit for arbitrary state preparation to encode $\mathbf{b}$ into $|\mathbf{b}\rangle$.

 A significant difference between HHL complexity and CG complexity is in error definition. In CG, $\varepsilon_c$ is defined as the energy norm of initial and final errors \cite{shewchuk1994introduction} (details in appendix) while $\varepsilon_h$ is defined as the two-norm of the difference between the HHL solution state $|\Tilde{\mathbf{x}}\rangle$ and the exact solution state $|\mathbf{x}\rangle = A^{-1}|\mathbf{b}\rangle$, denoted as $\varepsilon_h = ||\Tilde{\mathbf{x}}\rangle-|\mathbf{x}\rangle||_2$ \cite{dervovic2018quantum}. It's crucial to note that $\varepsilon_h$ does not signify the error between two classically readable solution vectors. A zero HHL error $\varepsilon_h$ (if achieved) would imply that HHL has `constructed' the exact quantum state representing $A^{-1}|\mathbf{b}\rangle$ and not the exact classical solution. 

If we analyze the runtime complexity of HHL without considering $T_b$, we observe that HHL exhibits exponential speedup with respect to the system size $N$, quadratic slowdown in sparsity $s$, and the condition number $\kappa$ \cite{dervovic2018quantum} in comparison to CG. Advancements in quantum linear system-solving methods have further widened this perceived quantum speedup from CG \cite{dalzell2023quantum}. However, these runtime complexities do not capture the time required to obtain a classically readable and usable solution vector $\mathbf{x}$. The HHL algorithm is only suitable for problems where the user is interested in measuring a few values of the form  $\langle\mathbf{x}|M|\mathbf{x}\rangle$\footnote{$M$ is a linear operator and HHL considers that we are interested in the expectation value of $M$ acting on $\widehat{x}$, the solution vector.} from the solution state $|\mathbf{x} \rangle$. 

In the DCPF (and subsequently QPF) problem, our interest lies in knowing all the entries of the vector, $\mathbf{x}$, representing bus angles. Therefore, the next subsection presents considerations of the complexity associated with quantum measurements or readout to obtain a classical description $\widehat{\mathbf{x}}$ of the HHL solution $|\widetilde{\mathbf{x}}\rangle$.

\subsection{Reading the Result: Measurement Complexity}
Upon solving a linear system using quantum computers, a classically readable vector $\widehat{\mathbf{x}}$ can only be obtained by measuring the quantum state solution $|\Tilde{\mathbf{x}}\rangle$. These quantum measurements are probabilistic in nature and the probability of an outcome is governed by Born's rule \cite{kockum2023lecture}. The process of obtaining a full classical description of the given quantum state is called quantum tomography \cite{abhijith2022quantum}. To perform quantum tomography for obtaining the bus angle vector estimate $\widehat{\mathbf{x}}$, access to multiple copies of corresponding HHL solution $|\Tilde{\mathbf{x}}\rangle$ is required. Further, the properties of QPF dictate that independent and identical measurements be performed on each copy of an HHL solution $|\Tilde{\mathbf{x}}\rangle$.
To obtain one copy of $|\Tilde{\mathbf{x}}\rangle$, linear system must be solved using HHL once. Therefore, to obtain the end-to-end complexity of solving QPF using HHL, we need to know how many copies of $|\Tilde{\mathbf{x}}\rangle$ are needed for readout to obtain a classical description of the solution $\widehat{\mathbf{x}}$. Using state-of-art results by considering that we have access to a unitary\footnote{We can consider access to a unitary because $|\Tilde{\mathbf{x}}\rangle$ is the output of a quantum algorithm \cite{van2023quantum}. Further, with this consideration, we require fewer copies of $|\Tilde{\mathbf{x}}\rangle$, thus obtaining a lower overall complexity.} (also the inverse of it) that prepares the state, we require $\Theta(\poly(N/\varepsilon_t))$ copies of $|\Tilde{\mathbf{x}}\rangle$ with $\varepsilon_t$ defined using $L_2$-norm. Here, $\Theta(\cdot)$ refers to simultaneous lower and upper bounds on asymptotic query complexity. Note that $\varepsilon_t \in [0,1]$ dictates the distance between a quantum state $|\Tilde{\mathbf{x}}\rangle$ and its classical description $\widehat{\mathbf{x}}$, obtained through multiple readouts. Considering that $|\mathbf{x}\rangle = A^{-1}|\mathbf{b}\rangle$ represents the exact solution and is analogous to $\mathbf{x}^\star$ in the classical CG setting, the net error in solving QPF is a combination of HHL errors $\varepsilon_q$ and $\varepsilon_t$. Discussion on the exact expression of QPF error and comparison with classical error $\varepsilon_c$ is omitted considering that both CG and HHL achieve final errors within an acceptable limit and error does not scale with system size $N$.

\begin{table*}[t]
    \centering
        \caption{QPF Solving Complexity Comparison in $\mathcal{O}(\cdot)$}
   \bgroup
\def\arraystretch{1.3}
    \begin{tabular}{c|c|c|c|c}
      Algorithm   & Solving Complexity & E2E Complexity & Optimistic E2E Complexity$^\#$ & with $\kappa = N^\beta$  \\
  \hline
           CG \cite{shewchuk1994introduction}     & $sN\sqrt{\kappa} \log(1/\varepsilon_c)$ & $Ns + sN\sqrt{\kappa} \log(1/\varepsilon_c) + N$ &  $s\sqrt{\kappa}N\log(N) \log(1/\varepsilon)$ &  $sN^{1+0.5\beta}\log(N)\log(1/\varepsilon)$  \\
           \hline
       HHL\cite{harrow2009quantum}          & $s^2\kappa^2\log(N) (1/\varepsilon)$ & $N/\varepsilon \Big (N\kappa + \log(N) \kappa^2 s^2/\varepsilon \Big)$  & $s^2\kappa^2N\log(N) (1/\varepsilon^2)$ & $s^2N^{1+2\beta}\log(N) (1/\varepsilon^2)$   \\
                VTAA-HHL \cite{ambainis2010variable}     & $s^2\log(N)\kappa \log^3(\kappa) (1/\varepsilon)$ & $N/\varepsilon \Big (N\kappa + \log(N) \kappa\log^3(\kappa) s^2/\varepsilon \Big)$ & $s^2N\log(N)\kappa \log^3(\kappa) (1/\varepsilon^2)$ & $s^2\beta N^{1+\beta} \log^4(N) (1/\varepsilon^2)$ \\
                Lowerbound \cite{harrow2009quantum, morales2024quantum}&  $\kappa \log(1/\varepsilon)$ & -- &  $\kappa N \log(1/\varepsilon)/\varepsilon $ & $ N^{1 +\beta}(\log(1/\varepsilon)/\varepsilon)$ \\
            \hline
            \multicolumn{5}{l}{$^\#$ Considering QRAM is available and $T_b= \log(N)$; $\quad$ E2E: End-to-End; $\quad$ Discussion on $\varepsilon_c$ is included in appendix}
    \end{tabular}
    \egroup
    \vspace{-2em}
    \label{tab:complexity}
\end{table*}

\subsection{End-to-End Complexity}
Now we analyze the end-to-end complexity of solving QPF. Starting from a state $|\mathbf{b} \rangle$, the depth of the HHL circuit required to prepare the solution $| \mathbf{\tilde{x}} \rangle$ with error $\varepsilon$ is $T_s = \mathcal{O}( \log(N) \kappa s^2/\varepsilon)$. This circuit only has a small success probability, therefore, the algorithm requires amplitude amplification with $\mathcal{O}(\kappa)$ repetitions of this circuit to succeed with high probability. Let $T_r$ be the complexity of readout with error $\varepsilon$. Assuming state-of-the-art quantum tomography methods, this complexity has a minimum complexity of $\mathcal{O}(N/\varepsilon)$ \cite{van2023quantum}. Let $T_b$ be the complexity of preparing the state $|\mathbf{b} \rangle$. Assuming no access to QRAM, this complexity cannot be lower than $\mathcal{O}(N).$ If we have a QRAM already loaded with this state, then this complexity becomes $\mathcal{O}(\log(N)).$ 

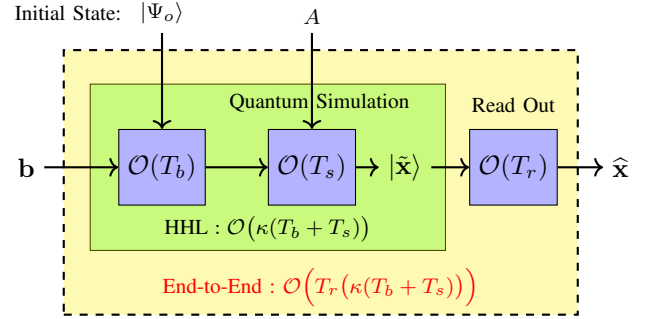
\begin{figure}[t]
    \centering
   \begin{tikzpicture}
    \node [draw, rectangle, minimum width=4.7cm, minimum height=2.2cm,fill=green!30] (p) at (0,0) {};
      \node [draw, dashed,thick,rectangle, minimum width=6.8cm, minimum height=3.5cm,fill=yellow,fill opacity=0.3] (p1) at (0.7,-0.2) {};
    \node [left=6mm of p] (in) {$\mathbf{b}$};
    \node [draw, rectangle, minimum width=1cm, minimum height=1cm, align=center,fill=blue!30] (b) at (-1.4,0) {$\mathcal{O}(T_b)$};
      \node [above=0.5in of b] (b1) {\footnotesize \textcolor{black}{$|\Psi_o\rangle$}};
       \node [left=0mm of b1] (b2) {\footnotesize \textcolor{black}{Initial State:}};
    \node [draw, rectangle, minimum width=1cm, minimum height=1cm, align=center, right=1mm of b,fill=blue!30] (A) at (-0.1,0) {$\mathcal{O}(T_s)$};
      \node [right=3mm of A] (out) {$|\Tilde{{\mathbf{x}}}\rangle$};
       \node [above=1mm of A,xshift=1mm] (A1)  {\footnotesize \textcolor{black}{Quantum Simulation}};
       \node [above=0.5in of A] (a1) {\footnotesize \textcolor{black}{$A$}};
      \node [draw, rectangle, minimum width=1cm, minimum height=1cm, align=center, right=5mm of out,fill=blue!30] (R){$\mathcal{O}(T_r)$};
        \node [above=1mmof R] (R1) {\footnotesize \textcolor{black}{Read Out}};
         \node [right=6mmof R] (R2) {$\widehat{\mathbf{x}}$};
       \node [above=0.01mm of p.south] (A2) {\footnotesize  {$\text{HHL}: \mathcal{O}\big(\kappa(T_b+T_s)\big)$}};
     \node [above=0mm of p1.south] (o) {\footnotesize \textcolor{red}{End-to-End : $\mathcal{O}\Big(T_r\big(\kappa(T_b+T_s)\big)\Big)$}};
    \draw [->,thick] (in) -- (b.west);
    \draw [->,thick] (b.east) -- (A.west);
    \draw [->,thick] (A.east) -- (out);
   \draw [->,thick] (out.east) -- (R);
   \draw [->,thick] (R.east) -- (R2);
    \draw [<-,thick] (b.north) -- (b1);
    \draw [<-,thick] (A.north) -- (a1);
\end{tikzpicture}
    \caption{Composition of end-to-end complexity of solving QPF using HHL. Here, $T_b$ represents state preparation runtime cost (in terms of circuit depth), $T_s$ is the cost of a single quantum simulation to solve linear system, and $T_r$ represents the number of copies of $|\Tilde{{\mathbf{x}}}\rangle$ needed to obtain $\widehat{\mathbf{x}}$ with a given error. With QRAM $T_b = \log(N)$, the end-to-end complexity is $\mathcal{O}\big( N\log(N) \kappa^2 s^2/\varepsilon^2 \big)$, and if QRAM is not available, than arbitrary state preparation complexity $T_b = N$ needs to be considered, which leads to end-to-end complexity of $\mathcal{O} \big( N^2\kappa/\varepsilon + N\log(N) \kappa^2 s^2/\varepsilon  \big )$ for complete $N$ state readout. \textcolor{black}{Readers can refer to \cite{harrow2009quantum,stepbystepHHL} for HHL's quantum circuit and Figure 1 of \cite{saevarsson2022quantum} for quantum circuits for solving QPF.}}
    \label{fig:sequence}
\end{figure}

Combining all of these complexities, the end-to-end complexity of QPF becomes $\mathcal{O}(T_r(\kappa(T_b +T_s)))$. This is pictorially represented in Figure~\ref{fig:sequence}. If we do not have QRAM access, optimistic estimates for this complexity become
\begin{equation}
\mathcal{O}\left(\frac{N}{\varepsilon}\kappa\left(N + \frac{\log(N) \kappa s^2}{\varepsilon}\right)\right).
\end{equation}
Assuming access to a QRAM pre-loaded with $| \mathbf{b} \rangle$, the most optimistic complexity estimate becomes
\begin{equation}
\mathcal{O}\left(\frac{N}{\varepsilon}\kappa\left( \log(N) + \frac{\log(N) \kappa s^2}{\varepsilon} \right)\right).
\end{equation}
 
As $\kappa \gg 1$ as shown in Figure \ref{fig:system_details} (center), the leading order, asymptotic complexity (considering the best case for PQA in QPF) is
\begin{align}\label{eq:e2e_qrm}
    \mathcal{O}\Big( N\log(N) \kappa^2 s^2/\varepsilon^2 \Big).
\end{align}

It is important to understand that \eqref{eq:e2e_qrm} does not represent the actual runtime of solving DCPF using the HHL algorithm. Rather, it gives the asymptotic scaling of this runtime with the parameters of the problem. This scaling is independent of the specific hardware and algorithmic implementations used and the exact runtime will depend on those factors.

\section{Discussion and Conclusion}
This section presents a comparative discussion between CG and end-to-end quantum complexity from a PQA viewpoint. In Table \ref{tab:complexity}, end-to-end complexities of quantum methods are compared with CG. 

 {Using the same methodology we also report the end-to-end complexities of a more advanced quantum system solver, namely HHL with Variable Time Amplitude Amplification (VTAA-HHL), in Table \ref{tab:complexity}. VTAA-HHL was introduced by Ambainis \cite{ambainis2010variable}. VTAA-HHL, combines HHL with VTAA, which is a method to boost the success probability of the HHL algorithm with improved complexity. As discussed before, the HHL circuit fails with probability $O(1/\kappa^2)$. In the vanilla version of HHL this is boosted to a constant success probability by using the amplitude amplification technique. Unlike vanilla amplitude amplification, which assumes a uniform runtime and applies a fixed number of Grover iterations, VTAA adapts the amplification process to selectively boost the success amplitude leading to a linear system solver that runs in time linear in $\kappa$ .  }

 {In final row of Table \ref{tab:complexity}, we also compute the theoretically optimal end-to-end complexity of a quantum linear system solver running on a fault tolerant quantum computer. This bound is derived from the known complexity lower bound for this problem obtained in the original HHL paper. This calculation necessarily lower bounds the solving complexity of all modern quantum solvers \cite{morales2024quantum}. We see that despite assuming this highly optimistic scenario with a theoretically optimal quantum solver and $O(\log(N))$ data loading overhead, the end-to-end complexity of the quantum solver is still worse than that of CG.}

\begin{figure*}[t]
    \centering
    \includegraphics[width=\textwidth]{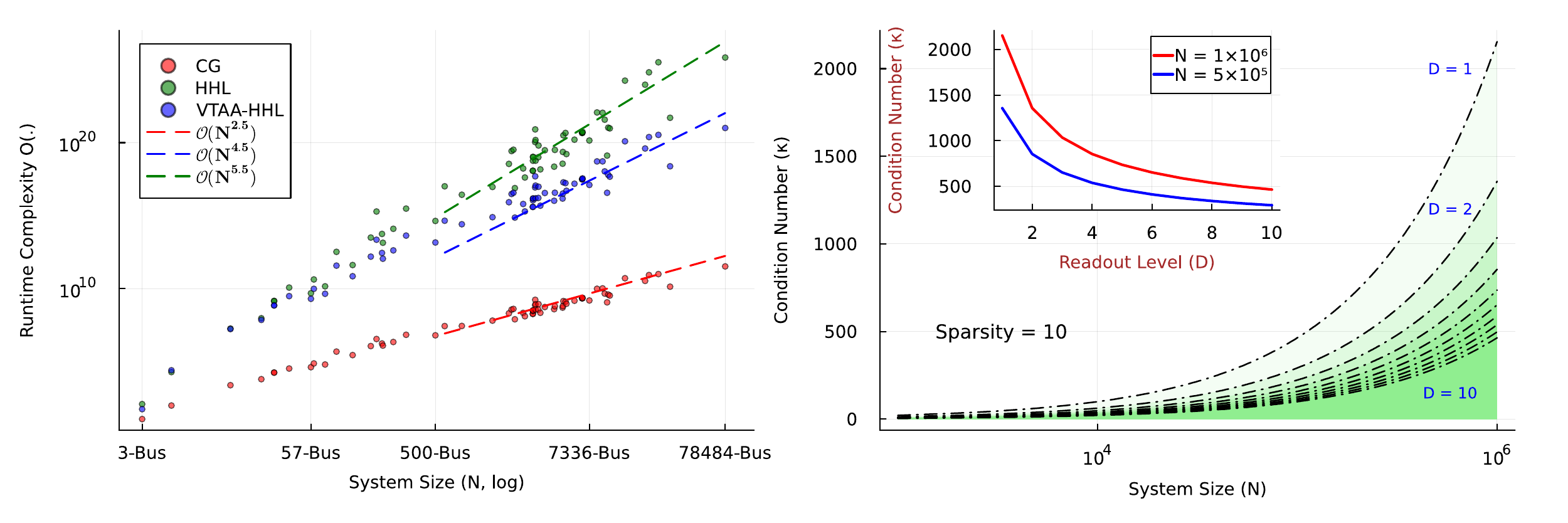}
    \caption{Runtime complexity comparison and potential PQA parameter range. \textbf{Left:} Complexity comparison between CG, HHL-Optimistic, and VTAA-HHL-Optimistic for power networks from \texttt{PGLib}\cite{pglib}. The fit beyond the 500-Bus system indicates that the approximate end-to-end runtime complexity of CG is $\mathcal{O}(N^{2.5})$, which is significantly lower than HHL's $\mathcal{O}(N^{5.5})$ and even VTAA-HHL's $\mathcal{O}(N^{4.5})$. Therefore, there is no practical quantum advantage in solving DCPF for these systems. \textbf{Right:} Upper bound of condition number ($\kappa$) values for achieving PQA using HHL complexity \eqref{eq:e2e_qrm} at different readout levels for system size $N$. Dotted black lines indicate different readout levels and $D=10$ means that only 10 classical values need to be extracted from the solution state $|\mathbf{x} \rangle$. Below these curves, the runtime complexity of CG is worse than that of HHL and there is a possibility of PQA. Importantly, for the \texttt{PGLib} systems, we cannot achieve any quantum speedup even for a single readout without reducing the condition number. It is clear from Table \ref{tab:complexity} and related discussion that PQA can only be achieved if the condition number is reduced using preconditioning and readout requirements are limited to a fraction of the state vector length. 
    Here, complexity dependence on error is ignored as errors do not scale with system size. Also, machine overheads are considered insignificant.}
    \label{fig:conclusion}
\end{figure*}

We find that an additional $\log(N)$ factor is required for CG complexity to make the error guarantees of CG and HHL comparable (see appendix for more details). Due to state preparation, readout (tomography) complexity, and condition number scaling, quantum algorithms have higher complexity than CG. This negates any chance of PQA in DCPF in this setting unless fundamental algorithmic barriers are overcome. Further, this end-to-end complexity comparison can directly be extended to the FDLF formulation presented in \cite{saevarsson2022quantum}. Even with the presence of QRAM, quantum computing runtime will be asymptotically worse in system size for FDLF compared to CG, even if FDLF takes $\mathcal{O}(\log(N))$ iterations to converge classically.\footnote{With linear system solving iterations growing as $\mathcal{O}(\log(N))$, FDLF can be solved classically with complexity $\mathcal{O}(N\log(N)\sqrt{\kappa}s\log(1/\varepsilon))$, which is better in terms of error, sparsity, and condition number from HHL complexity of the same with QRAM.} Therefore, practical quantum advantage does not exist for solving power flow problems (both DCLF and FDLF), primarily because of the requirement of reading out the complete solution vector and because of the complexity of encoding the injection vector into a quantum state. Figure \ref{fig:conclusion} (left) illustrates the runtime complexity comparisons using actual power network data (as shown in Figure \ref{fig:system_details}) and highlights that CG runtime complexity scales $\sim N^{2.5}$, which is significantly lower compared to HHL's runtime complexity growth $\sim N^{5.5}$ and VTAA-HHL's $\sim N^{4.5}$. 

One major computational bottleneck in the quantum approach is readout complexity, which scales linearly as $O(N).$ This can be reduced for some specific power flow application if only $D$ components of the solution vector need to be read instead of all  $N$ components. We call this the readout level of the problem and assume that $D \ll N$. This can reduce the readout complexity to $O(D).$ From our earlier complexity analysis, we formulate a relation between the various parameters in the problem ($D,s,\kappa,\ldots$) such that PQA becomes possible. As complexities do not represent exact runtimes, for there to be PQA, the complexity of CG must be greater (e.g., a factor of $N^\alpha$ where $\alpha>0$) than that of a quantum method. For the optimal HHL complexity from Table \ref{tab:complexity} to be substantially less than the complexity of CG, $D \kappa^{3/2} \ll N s^{-1}$ is required. Similarly, in the optimistic scenario with the VTAA-HHL algorithm, we get $D\sqrt{\kappa} \log^3(\kappa) \ll Ns^{-1}.$ These relations act as a general rule to check for the existence of PQA in these problems.

Even in a regime where this rule of thumb does not hold, quantum computers may be able to solve specific DCPF instances faster than CG for small systems. But this doesn't imply a fundamental quantum speedup. Moreover near-term fault tolerant quantum computers are expected to have significant error-correction overheads \cite{babbush2021focus,noh2022low}. Thus, for PQA these overheads should be offset by a significantly favorable complexity scaling for the quantum algorithm. This can further narrow the window of parameters for which PQA can exist.

 Figure~\ref{fig:conclusion} (right) illustrates the interplay between condition number $\kappa$, system size $N$, and readout level $D$ for attaining PQA at constant sparsity. 
 Notably, higher readout levels demand lower condition numbers for PQA. In a power system, it is well-known that the number of branches reaching or violating desired limits is only a fraction of the total number of branches. Thus, a problem to identify the power flow on candidate violating branches can be cast as a linear system with very low readout level requirements. However, for PQA, this system would still require preconditioning to reduce the condition number. Therefore, exploring techniques to mitigate condition number growth is essential for achieving PQA. Most importantly, the conditions in Figure~\ref{fig:conclusion} highlight the possibility of having PQA with at most sub-quadratic speedup. For near-exponential speedup, both the condition number and readout requirements should ideally not scale with the size of the system.



 In conclusion, our end-to-end analysis of QPF complexity reveals a nuanced picture of PQA in power flow calculations. Although initial claims of exponential speedups proved illusory, this research pinpoints a narrow window where PQA might be achievable. For DCPF and FDLF problems, our findings demonstrate the absence of PQA under typical network parameters and problem specifications. However, for DCPF-type problems with specific condition number values and readout requirements, a sub-quadratic speedup may be attainable. This highlights the need for further research into tailored QPF algorithms and problem settings to unlock the true potential of quantum computing for power grid analysis. This work provides a critical step towards demystifying the hype surrounding QPF and understanding the requirements for future breakthroughs in this field.

\bibliographystyle{IEEEtran}
\bibliography{main}

\appendix




\section*{A. Brief Explanation of the CG Algorithm}
CG iteratively finds the minimum of $f(\mathbf{x}) =  \frac12 \mathbf{x}^TA\mathbf{x} - \mathbf{b}^T\mathbf{x} + c$, which is a solution of $A \mathbf{x}=  \mathbf{b}$ if $A$ is positive definite. The main idea of CG is to generate vectors at each iteration that are conjugate (or A-orthogonal) to all previous iteration vectors and only require the $i$-th conjugate vector for generating the $i+1$-th conjugate vector\cite{shewchuk1994introduction}. Using the conjugate property of these vectors (called direction vectors), CG is shown to require at most $N$ iterations to converge to the solution of $A\mathbf{x} = \mathbf{b}$, from any starting point $\mathbf{x}^o \in \mathbb{R}^N$. However, the floating point round-off errors and cancellation errors prevent using CG to find an exact solution. Also, for DCPF problems, we are interested in finding a solution while accepting $\texttt{error} \leq \texttt{Measuring Device Precision}$.
 In performing these conjugate vector calculations and updated states in CG, \textit{matrix-vector} multiplication becomes the dominant operation \cite{shewchuk1994introduction}. For a sparse matrix with at max $s$ non-zero elements in a row, one \textit{matrix-vector multiplications} will have complexity $\mathcal{O}(Ns)$.

 The second part of CG complexity depends on the number of these \textit{matrix-vector} multiplications to find the solution $\mathbf{x}^\star$ with an acceptable error $\varepsilon_c$. This means that the number of iterations required can be calculated based on error convergence after $i$-iterations, i.e., the relationship between initial- and post-$i$-th iteration error. Let initial error vector $\mathbf{e}^o = \mathbf{x}^\star - \mathbf{x}^o $, and after $i$ iterations, $\mathbf{e}^i = \mathbf{x}^\star - \mathbf{x}^i $ then error convergence will be $\|\mathbf{e}^i\|_{A}\leq \varepsilon_c\|\mathbf{e}^o\|_A$. Here, $\|\mathbf{e}\|^2_{A} = \mathbf{e}^TA\mathbf{e}$ is sometimes referred to as energy-norm \cite{shewchuk1994introduction} and $\varepsilon_c$ is the error constant that determines how close CG convergence is to the actual solution with respect to the initial error \big($\varepsilon_c \geq \|\mathbf{e}^i\|_{A}\big /\|\mathbf{e}^o\|_{A}$ \big). Further, using convergence analysis from \cite{shewchuk1994introduction}, the maximum number of iterations needed to achieve $\varepsilon_c$ is given as
  \begin{align*}\label{eq:iter}
      i\leq \bigg\lceil \frac{1}{2}\sqrt{\kappa}\log(2/{\varepsilon_c})\bigg \rceil
  \end{align*}
The square root dependence of CG runtime complexity on condition number $\kappa$ is beneficial because a power system network matrix $A$ tends to have a high condition number and approximately quadratic growth of $\kappa$ with respect to system size $N$ for \texttt{PGLib} networks. The feature of power system matrices having high condition numbers affects the convergence of CG as a high condition number and implies a higher eigenvalue spread, meaning a large number of iterations are required to converge \cite{shewchuk1994introduction}. Preconditioning of the linear system $A\mathbf{x} = \mathbf{b}$ can solve this problem. However, finding a computationally cheap and stable preconditioner is a non-trivial problem \cite{shewchuk1994introduction}.

\begin{figure}[h]
    \centering
    \includegraphics[width=0.4\textwidth]{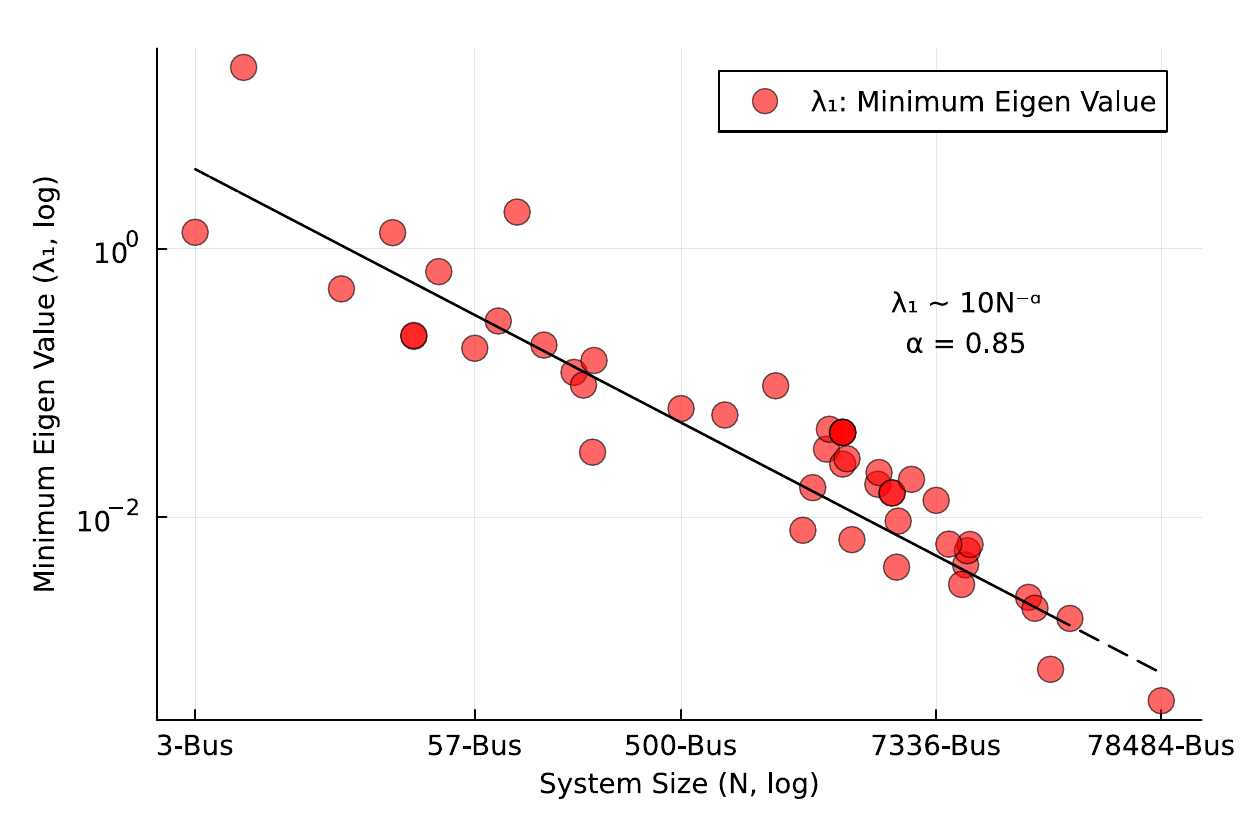}
    \caption{Decay of minimum eigenvalue with system size for \texttt{PGLib} systems \cite{pglib} (except \textit{PEGASE} systems).}
    \label{fig:eigen}
\end{figure}

For comparing the complexity of CG with HHL-QPF, it's crucial to establish the relationship between $\varepsilon_c$ (error defined using the energy norm) and $\varepsilon$ (error defined using the Euclidean norm). If $A$ is positive semidefinite (PSD) and $\lambda_1 < \dots < \lambda_N$ represents the ordered set of eigenvalues of $A$, then $\lambda_1\varepsilon \leq \varepsilon_c$. Thus, running CG until the energy norm of the error is less than $\varepsilon/\lambda_1$ guarantees that the $L_2$ error in the solution is less than $\varepsilon$. This unifies the error guarantees of CG and HHL and brings their complexities onto the same footing. Additionally, as shown in Figure \ref{fig:eigen}, for \texttt{PGLib} systems, the smallest eigenvalue grows approximately as $N^{x}$ with $N$ being the number of buses. Consequently, the CG complexity in Table \ref{tab:complexity} will include a $\log(N)$ term when $\varepsilon_c$ with $\varepsilon$ is substituted in the end-to-end complexity.

\section*{B. Brief Explanation of HHL}\label{app:hhl}
The algorithm starts by encoding load vector $|b\rangle$ information onto an $n=\log(N)$ qubit register using a unitary operation $U_b$, such that $U_b|0\rangle = |\mathbf{b}\rangle$. This action signifies that the initial value register is now in a state containing the information from $|\mathbf{b}\rangle$. Subsequently, the quantum phase estimation subroutine is employed to encode the eigenvalues of matrix $A$ into an $n_c$-qubit register. The number of qubits ($n_c$) determines the accuracy with which the phase values, i.e., eigenvalues, are stored. The next step involves applying controlled rotation to an ancilla register (one qubit) based on the eigenvalues stored in the $n_c$-register. This rotation uses the principle that for a Hermitian $A$, the eigenvalues of $A^{-1}$ are the inverses of the eigenvalues of $A$, using the spectral decomposition theorem. In other words, if $A = Q \Lambda Q^\dagger$ with $Q$ as the eigenvector matrix and $\Lambda$ as the diagonal matrix with eigenvalue entries, then $A^{-1} = Q \Lambda^{-1} Q^\dagger$. Upon rotation, the probability of obtaining $|1\rangle$ in the ancilla register outcome, through measurement, is proportional to the eigenvalues of $A^{-1}$, i.e., the inverses of the eigenvalues of $A$. Then an inverse phase estimation is performed to return the $n_c$-qubit register to its base state $|0\rangle$, and finally, the ancilla is measured. If the ancilla measurement result is $|1\rangle$, then the algorithm succeeds and the solution is provided as a quantum state. Otherwise, the entire process starting from state preparation to ancilla measurement is repeated until the state $|1\rangle$ is achieved. The final quantum state $|\Tilde{\mathbf{x}}\rangle$ is proportional to the solution vector $|\mathbf{x}\rangle$. Readers can refer to \cite{harrow2009quantum,stepbystepHHL,abhijith2022quantum,kockum2023lecture} for details and an explanation of how HHL works.

\section*{C . Additional Caveats}

{(i) \textbf{Hot-starting}:} It is clear that if classical and quantum computing errors are less than a threshold, they have a lesser impact on comparative runtime performance analysis. However, by the definition of classical error, $\varepsilon_c$, it is clear that hot-starting or warm-starting a CG-based DCPF solving algorithm can lead to faster runtimes as discussed in the main section. In practical situations, we run multiple DCPFs with load variations within a range. Thus, warm-starting will benefit in reducing the overall time taken by the DCPF running over multiple instances. Although the idea of warm-starting quantum algorithms has been presented in literature \cite{truger2023warm}, potential benefits, if any, will require closer scrutiny and dedicated quantum linear system warm-starting methods.

{(ii) \textbf{Condition number limitations}:}  
The HHL algorithm (and other variants) are shown to be robust when the $A$ matrix is well conditioned, i.e., has a low condition number. However, in a power system, the $A$ matrix condition number grows approximately quadratically with the system size, as shown in Figure \ref{fig:system_details}(center). Authors in \cite{saevarsson2022quantum} discuss this limitation on condition number and suggest that preconditioning methods can be developed to artificially reduce the condition number of the $A$ matrix. However, there are a few limiting factors that will impact any envisioned benefit from preconditioning:
\begin{itemize}
    \item Preconditioner selection is not a trivial problem. Also, the complexity of preconditioning will modify the overall complexity of the linear system-solving process. To maintain a practical quantum advantage, preconditioning complexity must be less than that of solving the linear system itself. For example $A^{-1}$ is a preconditioner of $A$ as $A^{-1}A = \rm I$. However finding $A^{-1}$ is equivalent to solving the linear system $A\mathbf{x} = \mathbf{b}$. Therefore, it is difficult to establish if PQA will exist for power flow problems without analyzing specific preconditioning algorithms.
    \item Benefits of preconditioning will be available for both classical and quantum algorithms given that preconditioning complexity is acceptable as discussed prior. Also, note that HHL's runtime complexity dependence on condition number $\kappa$ is worse than that of CG's. This implies that condition number reduction will impact only the robustness of QPF methods and is unlikely to play any role in deciding PQA for QPF. 
\end{itemize}


\end{document}